\documentclass{aa}  
\usepackage{graphicx}
\usepackage{txfonts}
\newcommand{\teff}{\ensuremath{T_{\rm eff}}}
\newcommand{\lv}{local 20\,pc volume}
\newcommand{\bz}{\ensuremath{\langle B_z \rangle}}
\newcommand{\bs}{\ensuremath{\langle \vert B \vert \rangle}}
\begin{document} 

   \title{Highly sensitive search for magnetic fields in white dwarfs using broad-band circular polarimetry}

   \author{Andrei V. Berdyugin\inst{1}
           \and
           Vilppu Piirola\inst{1}
          \and
          Stefano Bagnulo\inst{2}
          \and
          John D. Landstreet\inst{2,3}
          \and
          Svetlana V. Berdyugina\inst{4}
          }
   \institute{Department of Physics and Astronomy, FI-20014 University of Turku, Finland;
              \email{andber@utu.fi {\rm and} piirola@utu.fi}
         \and
             Armagh Observatory \& Planetarium, College Hill, Armagh BT61 9DH, UK;
             \email{stefano.bagnulo@armagh.ac.uk}
         \and
         Department of Physics \& Astronomy, University of Western Ontario, London, Ontario, N6A 3K7, Canada;\\
         \email{jlandstr@uwo.ca}
         \and
         Leibniz-Institut f\"ur Sonnenphysik (KIS), Sch\"oneckstr 6, Freibirg, Germany
         \email{svetlana.berdyugina@leibniz-kis.de}
             }

   \date{Received September 15, 2021; accepted October 29, 2021}

  \abstract{
Circular polarisation measurements of white dwarfs of various ages and spectral types are useful to understand the origin and evolution of the magnetic field in degenerate stars. In the latest stages of white dwarf evolution, when stars are so cool that spectral lines are no longer formed in the normal H- or He-dominated atmospheres, magnetic fields can be probed only by means of circular polarimetry of the continuum. The study of the fields of featureless DC white dwarfs may reveal whether Ohmic decay acts on magnetic white dwarfs, or if magnetic fields continue to be generated even several billion years after white dwarf formation. Compared to spectropolarimetry, broad-band circular polarisation measurements have the advantage of reaching a higher accuracy in the continuum, with the potential of detecting magnetic fields as weak as a fraction of a MG in DC stars, if the telescope size is adequate for the star's magnitude. Here we present the results of a first (short) observing campaign with the DIPol-UF polarimeter, which we have used to measure broad-band circular polarisation of white dwarfs. Our observing run was in part aimed to fully characterise the instrument, and in part to study the relationship between magnetic field strength (when known from spectropolarimetry) and circular polarisation of the continuum. We also observed a small number of previously unexplored DC white dwarfs, and we present the discovery of two new magnetic white dwarfs of spectral class DC, probably the first discovery of this kind made with broad-band circular polarimetric techniques since the late 1970s. We also discuss the characteristics of our instrument, and predict the level of polarimetric accuracy that may be reached as a function of stellar magnitude, exposure time, and telescope size.
}

   \maketitle


\section{Introduction}
White dwarfs (WDs) are the end point of 90\,\% of stellar evolution. About 20\,\% of such stars possess strong magnetic fields, but if we exclude  stars younger than 0.5\,Gyr from
the statistics, this frequency rises to almost 30\,\% \citep{BagLan21}. The distribution of the field strength is approximately uniform per decade, in a range that varies from a few tens of KG to several hundred MG, and there are no obvious signs of Ohmic decay during WD cooling, at least until an age of 5\,Gyr \citep{LanBag19a,BagLan21}. The origin of these fields is not fully understood. Several hypotheses have been put forward: for instance, a field could be inherited from pre-WD evolution, or originate during the merging of a binary system, or it might be produced by a dynamo acting after crystallisation begins \citep{Isern2017}. None of the proposed mechanisms is able to explain all observed magnetic features of degenerate stars, and more than one channel of field formation may exist. 

Magnetic fields are primarily detected through the analysis of the Zeeman effect on the Stokes profiles of spectral lines produced in the atmospheres of almost pure H or He  found in most WDs. Zeeman splitting may be detected with high-resolution spectroscopy in DA, DB, and DZ WDs (with prominent H, He,  and sometimes Ca or Fe lines) that have  field strengths as low as 50\,kG.  More commonly, from low-resolution spectroscopy, the usual detection limit is of the order of 1\,MG. With spectropolarimetry, the detection threshold  is 1 -- 3 dex lower. Circular polarimetry is sensitive to the component of the magnetic field along the line of sight averaged over the stellar disk. For this quantity, usually called  mean longitudinal field (\bz), values as small as a few kG may be detected in DA and DB stars \citep{Aznar2004, LanBag19a}. 

The situation is different in featureless (DC) stars, in which a magnetic field may be detected only if its strength is sufficiently high to circularly polarise the continuum. Assuming that a 15\,MG field is needed to produce 1\,\% of polarisation \citep{BagLan20}, the weakest fields detected so far in DC WDs are of the order of several MG. However, it is in principle possible to firmly detect circular polarisation signals of the order of a few parts in  $10^{-4}$, or even smaller, and thus to detect magnetic fields in DC WDs as weak as a fraction of a MG (if such fields exist in DC stars). \citet{BagLan21} have suggested that the field strength may be roughly uniformly distributed per decade in a range between a few tens of kG and few hundred MG. Recalling that the magnetic frequency in older stars is of the order of 30\,\%, we conclude that with appropriate exposure time and telescope size, and with an instrument capable of delivering a polarimetric sensitivity of order $10^{-4}$, one would expect to detect a magnetic field in at least one out of five DC WDs.

Estimating the frequency of magnetic fields in DC WDs as a function of stellar age would be very useful to establish whether Ohmic decay operates during DC WD cooling, and if so, on which timescale; whether  the field strength starts to decrease with time after a certain age; and if there is an age at which none of the DC stars are found to be magnetic. These data would help to set strong constraints to the theories that try to explain the origin and evolution of magnetic fields in degenerate stars.

The circular polarization of the continuum may be measured with a photo-multiplier tube (PMT) or in imaging mode with a CCD detector, through narrow- or broad-band filters, or even in white light without any filter at all; or with spectropolarimetric techniques. Narrow- and broad-band circular polarisation (BBCP) measurements were extensively used in the 1970s \citep[e.g.][]{Kemetal70,AngLan70a,LanAng71} and the early 1980s \citep{LieSto80,Angetal81}, but have never been systematically exploited during the past four decades. However, spectropolarimetry of the continuum for surveys of DC WDs has been used by \citet{Putney97} and more recently, by \citet{BagLan20} and \citet{BagLan21} for a systematic survey of DC WDs of the \lv. 

Spectropolarimetry potentially has a much higher diagnostic content than BBCP because the detailed behaviour of the polarised spectrum with wavelength is dictated by the field strength, atmospheric chemistry, and morphology. Spectral resolution may be traded with the signal-to-noise ratio, S/N, via software, allowing one in principle to reach the same S/N as in BBCP after heavy re-binning, if this is necessary with faint stars. However, \citet{BagLan20} and \citet{BagLan21} have argued that the spectropolarimetry of the continuum is generally affected by higher instrumental and background polarisation than BBCP, making it effectively very difficult to assess the reliability of any signal smaller than $10^{-3}$, no matter how high the S/N is. BBCP techniques have been proved to reach a sensitivity better than $10^{-4}$ and should therefore be regarded as the technique of choice for the detection of the weakest fields in DCs stars, but also for larger surveys, to be followed up by spectropolarimetric monitoring once a magnetic WD has been identified. 

We have developed high-precision polarimeters for applications of broad-band linear polarimetry, ranging from studies of the minute polarisation produced by interstellar dust and the magnetic field in the solar neighbourhood to strongly interacting binary stars with a BH component. Our Dipol-2 polarimeter \citep{Piietal14} and the more recent version, DIPol-UF \citep{Piietal21}, achieve precision at the $10^{-5}$ level. We expect to reach a similar performance level in circular polarimetry for sufficiently bright stars. With this in mind, we initiated a pilot study to explore the capabilities of the instrument in circular polarimetry of magnetic WDs, and the prospects for applying the BBCP techniques in general. In this paper we discuss the results of our study, and report new discoveries of previously  unknown magnetic WDs.

\section{Observations and data reduction}
\subsection{Instrument description}
The observations were carried out at the 2.5\,m Nordic Optical Telescope 
(NOT, located at the Observatorio del Roque de los Muchachos, La Palma, Canary Islands) on three nights, 
2021 July 2-5, using the simultaneous three-colour ($B'V'R'$) polarimeter DIPol-UF. The polarimeter is
described in detail by \citet{Piietal21}.

The light enters DIPol-UF through the modulator, a B. Halle superachromatic 
quarter-wave plate (QWP) for circular polarisation measurements. The plate is
rotated to discrete positions with steps of 90$\degr$ for circular polarimetry, 
and it remains at a fixed angle while each exposure is in progress. The optical axis of the QWP is at 45$\degr$ from the polarisation planes of the analyser.
A plane-parallel calcite plate acts as a polarisation analyser
and splits the incident radiation into two orthogonally linearly polarised
rays (ordinary and extraordinary). With the image scale of the NOT, the o- and e-image separation is about 14.6".
Two dichroic beam splitters split the light into the three ($B'V'R'$)
spectral passbands, with equivalent wavelengths of 445, 540,
and 640\,nm, and a full width at half maximum, FWHM, of 114, 75, and 96\,nm, respectively (Fig.1). We emphasise that our $B'V'R'$ bands are not those of the Johnson-Cousins system, since broad overlapping passbands cannot be created by the dichroic beam splitters. Instead, we use sharp-cutoff filters with high peak transmission. The $V'$ band is somewhat narrower than the other bands due to the particular filter used for that band in addition to the dichroic beam splitters.

\begin{figure}
\centering
    \includegraphics[keepaspectratio, width = 1\linewidth]{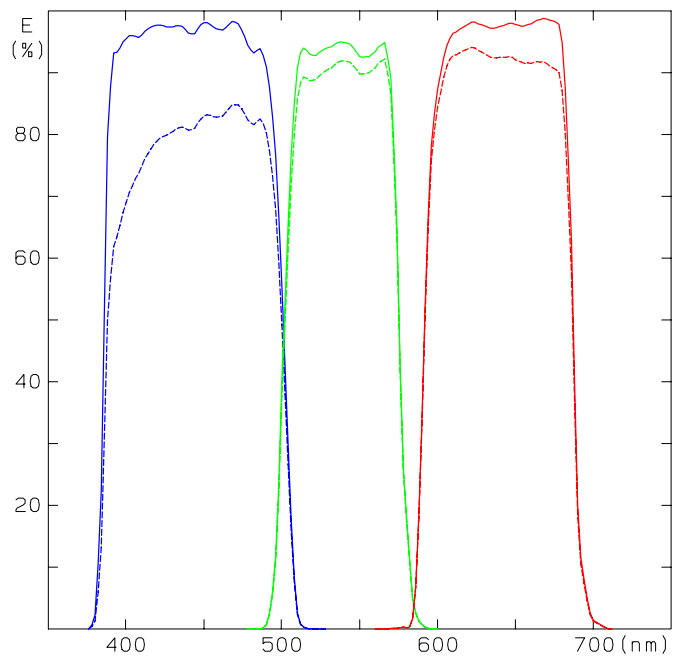}
    \caption{DIPol-UF passbands ($B'V'R'$) determined by the filters and dichroic beam splitters, 
    in the blue ($B'$), green ($V'$), and red ($R'$) spectral region (continuous lines). Dashed lines show the effect of the quantum efficiency of the CCD detectors taken into account. }
    \label{fig:passbands}
\end{figure}

DIPol-UF uses three ultrafast electron-multiplied (EM) Andor iXon  CCD cameras. 
The $V'$ and $R'$ cameras are identical iXon Ultra 897-EX models, while for the $B'$ 
passband, the iXon Ultra 897-BV model with the sensitivity optimised for the 
blue region is used. Faint targets are recorded in the EM amplifier mode with 
the EM gain value in the range of 5–20. In the EM mode, the camera can achieve 
single-photon sensitivity owing to the elimination of the readout noise, which 
significantly increases the S/N. For bright targets, conventional 
amplifiers with higher dynamical range are used in combination with a defocusing 
technique. The minimum number of the optical elements, high throughput, negligible 
systematic effects, and direct optical elimination of sky polarisation make 
DiPol-UF a very efficient instrument for broad-band polarimetry. The sensitivity of 
DIPol-UF in measuring polarisation is limited in practice by photon statistics 
and depends on the number of registered photons and the achieved S/N.

Each two successive exposures with orthogonal orientations of the QWP provide
one circular polarisation measurement. 
The uncertainty of circular polarisation is the inverse of the S/N of the flux measured on both beams accumulated over all exposures. For a general formulation, see for example \citet{patat2006}.
For targets in the magnitude range
15 $< V <$ 17, typically 20-40 sec individual exposure times were used. The delay 
between the exposures due to camera readout and QWP rotation is $< 0.3$\,sec and is
therefore negligible. In a typical nightly monitoring interval of 0.5-1.5 hours, 
the number of individual measurements for one object is in the range 32 - 96. 
These data provide reliable statistical error estimates.

\subsection{Target selection}

\begin{table*}
  \caption{\label{Tab_Programme} Programme stars and their main physical features.}
\begin{tabular}{llrrrcccl}
\hline\hline
\multicolumn{2}{c}{STAR}               &  $G$ &\multicolumn{1}{c}{$d$}& \teff & log $g$ &    $M$      & Age  & Previous observations \\
               &                       &      &(pc)  & (K)   &  c.g.s. & ($M_\odot$) &(Gyr) & \\
\hline
WD\,1309$+$853  & G\,256-7               & 15.8 & 16.5 &  5300 &8.12&0.66&5.46& DA MWD with $\bs \simeq 5.4 $\,MG (7)\\ 
WD\,1350$-$090  & GJ 3814                & 14.6 & 19.7 &  9580 &8.13&0.68&0.81& DA MWD with $\bs \simeq 0.45$\,MG (5,15)\\
WD\,1503$-$070  & GD 175                 & 15.8 & 20.3 &  6640 &8.00&0.59&1.80  & DA MWD with $\bz \sim 2$\,MG (12)   \\
WD\,1556$+$044  & PM J15589+0417         & 16.0 & 22.5 &  6685 &8.43&0.85&3.65& DC (He) never observed \\ 
WD\,1647$+$591  & DN Dra                 & 12.3 & 10.9 & 12738 &8.24&0.76&0.45& Non magnetic DA WD (13)\\ 
WD\,1748$+$708  & G 240-72               & 13.8 &  6.2 &  5570 &8.34&0.79&5.86& Known strongly polarised DQ WD (2,4,9)\\
WD\,1831$+$197  & G 184-12               & 16.3 & 39.3 &  7305 &8.04&0.60&1.57& DQ (He) no polarisation detected (8)\\  
WD\,1856$+$534  & LP 141-14              & 17.0 & 24.8 &  4430&7.69&0.40&4.34& DC (He) never observed \\
WD\,1900$+$705  &Grw\,+70$^\circ$\,8247  & 13.3 & 12.9 & 11880 &8.54&0.93&0.92& Known strongly polarised DA WD (1)\\
WD\,1953$-$011  & LAWD 79                & 13.6 & 11.6 &  7868 &8.23&0.73&1.63& DA6 MWD ($\bz \sim 0.1$\,MG with spot) (6,10)\\
WD\,2011$+$063  & G 24-9                 & 15.7 & 22.9 &  6455 &8.10&0.63&2.34& DC (He) no polarisation detected (3,8)  \\%
WD\,2032$+$248  & HD 340611              & 11.6 & 14.8 & 20704 &8.03&0.64&0.06& Non magnetic DA WD (13)\\ 
WD\,2049$-$253  & UCAC4 325-215293       & 16.0 & 18.0 &  4895 &7.84&0.48&4.4 & DC MWD (14)\\ 
WD\,2049$-$222  & LP 872-48              & 15.0 & 20.3 &  8164 &8.33&0.79&1.97& DC (He) never observed \\  
WD\,2117$+$539  & EGGR 378               & 12.4 & 17.3 & 15250 &7.94&0.58&0.17& Non magnetic DA WD (13)\\
WD\,2254$+$076  & G 28-27                & 17.1 & 45.5 & 12847 &9.26&1.29&$\sim 1.8$& DA MWD with $\bs \simeq 16$\,MG (11)\\
\hline
\end{tabular}
\tablefoot{
Key to references: 
1: \citet{Kemetal70}; 
2: \citet{Angetal74};
3: \citet{Angetal81}; 
4: \citet{West89};
5: \citet{SchSmi94}; 
6: \citet{SchSmi95}; 
7: \citet{Putney95};
8: \citet{Putney97}; 
9: \citet{BerPii99};
10: \citet{Valetal08};
11: \citet{Kuletal09};
12: \citet{Giaetal11};
13: \citet{BagLan18}, 
14: \citet{BagLan20}; 
15: \citet{BagLan21}. 
}
\end{table*}

Our target list comprises some WDs that are well studied in circular polarisation, and some WDs that have never been observed before. 
Well-studied targets include (1) WDs for which highly sensitive spectropolarimetric measurements failed to detect any magnetic field, and which  may therefore be considered as non-magnetic standard WDs, (2) WDs that have relatively weak fields (from hundreds to several thousand kG), and (3) WDs that are known to exhibit a signal of BBCP.  In addition to this group of 11 stars, we observed 5 DC WDs that were never observed before or in which we wished to confirm a previous detection. Our programme stars are listed in Table~\ref{Tab_Programme}, together with a summary of each star's main features.

\subsection{Data reduction}

\begin{table*}
\caption{\label{Tab_Log_Std} Observing log of bright non-polarised standard stars.}
\begin{center}
\begin{tabular}{lrcccrr@{$\pm$}lr@{$\pm$}lr@{$\pm$}l}
\hline\hline
\multicolumn{1}{c}{STAR} & $G$ & DATE & UT & JD -- & Exp. & \multicolumn{6}{c}{$V/I$ (\%)} \\
                         &     &yyyy-mm-dd & hh:mm &2400000 &\multicolumn{1}{c}{(s)} &
\multicolumn{2}{c}{$B'$} &\multicolumn{2}{c}{$V'$} &\multicolumn{2}{c}{$R'$} \\
\hline
HD\,115043  & 6.6 & 2021-07-02 & 21:19 & 59398.389  & 1280 &$-0.000 $& 0.001 &$-0.001$& 0.001&$-0.000 $& 0.001 \\
HD\,122676  & 6.9 & 2021-07-03 & 21:11 & 59399.383  & 1024 &$ 0.001 $& 0.001 &$ 0.001$& 0.001&$-0.000 $& 0.001 \\
HD\,124694  & 7.0 & 2021-07-04 & 21:19 & 59400.389  & 1152 &$-0.003 $& 0.001 &$-0.000$& 0.001&$ 0.001 $& 0.001 \\
HD\,211476  & 6.9 & 2021-07-03 & 05:00 & 59398.708  & 1600 &$ 0.002 $& 0.001 &$ 0.001$& 0.001&$-0.000 $& 0.001 \\
\hline
\end{tabular}
\end{center}
\end{table*}

After the standard CCD image reductions (bias, dark), the fluxes of the o- and 
e-images were extracted with an aperture-optimising photometric algorithm that minimises the standard error of the mean value of the circular polarisation calculated from
the individual measurements. The difference between the o- and e-beam intensities at the orthogonal orientations of the QWP (Sect. 2.1) yield one measurement of the Stokes parameter $V/I$, which is usually described as the net percentage circular polarisation. 
Depending on the seeing and star brightness, the optimum aperture 
radius was in the range 1.2" to 2.8", and the sky annulus thickness was 1.5" to 2.2".
To calculate the final value of $V/I$ from individual measurements, we applied the 2$\sigma$ weighting algorithm \citep{piirola2020a}, which gives a lower weight for points deviating by $>2\sigma$. Normally, 5-8 \% of the points deviated
by more than 2$\sigma$ and were given a weight $W<1$. The remaining points (92-95\,\%) were equally weighted ($W=1$). The algorithm helps to suppress the effects of cosmic rays, transient clouds, or occasional moments of poor seeing. 

\section{Interpretation of the observed circular polarisation}

It is important to realise that the presence of a magnetic field in a WD may lead to detectable broad-band circular polarisation through two quite different mechanisms. One mechanism is related to the continuum, and the other to the line opacity.  

First, the spectrum of the star may include continuum radiation, outside even of the wings of possible discrete spectral features, which is circularly (and perhaps linearly) polarised. 
These polarised continua are expected to be produced by the subtle dependence of the sources of continuum opacity in the WD atmosphere on the magnetic field, so that the continuous opacity at a given wavelength is greater in one sense of circular polarisation than in the other, and the emergent continuum radiation has a net polarisation. The continuous opacity in cool H- or He-dominated WD atmospheres is due to combinations of opacity produced by free electrons, neutral atoms, negative atomic ions, and molecular ions. The result is that continuum polarisation is expected to depend on composition, \teff, and on the presence or absence of metallic impurities.  A particularly simple treatment of this basic idea is described by \citet{Ship71}. 

An example of the presence of true continuum polarisation is shown in Fig.\,2 of \citet{PutJor95}. In this figure, it is clear that in addition to the circular polarisation spikes coinciding with the $\sigma$ components of H$\alpha$, there is a baseline of non-zero continuum circular polarisation, in spite of the absence of any absorption features other than the Zeeman triplet. This baseline continuum polarisation is produced by the continuous opacity effects mentioned above. 

To provide a first estimate of magnetic field strength from detected non-zero continuum circular polarisation, we often make use of a simple order-of-magnitude estimate of the value of the line-of-sight average field \bz\ that causes the observed polarisation. This estimate assumes that the observed continuum value of $V/I$ is very roughly proportional to \bz, with a proportionality constant of about 15\,MG per 1\,\% circular polarisation. The origin of this estimate and derivation of the proportionality constant are discussed in detail by \citet{BagLan20} and references therein. 

The resulting estimate of \bz\ can then be used to provide a (still less accurate) estimate, or at least a lower limit, of the mean surface field strength \bs\ by noting that \bs\ must be larger than its line-of-sight component ($\bs > \bz$), and  we generally find $\bs \geq 2.5 \bz$ from experiments in which these two field strength
values were computed for various dipolar and dipole-like magnetic field structures. 

A second mechanism that leads to non-zero circular polarisation in broad-band filter measurements arises in MWDs that have strong polarised spectral features, such as polarised Zeeman splitting in the Balmer lines. Two examples are shown by \citet{Putney97} for GD\,90 (with a field of $\bs \approx 9$\,MG) in her Fig.\,1{\it h} and for PG\,1658+441 (with a field of $\bs \approx 2.3$\,MG) in Fig.\,1{\it j}. For both these stars, it is clear that the Zeeman pattern of H$\alpha$ is practically symmetric, and so in a broad-band measurement, the two $\sigma$ components will contribute approximately equal numbers of polarised photons, which because of the opposite signs of $\sigma_+$ and $\sigma_{-}$ will lead (approximately) to no net polarised signal in the filter band. 

However, at shorter wavelengths, it is no longer obvious that polarised counts contributed by the blue and red polarised signals from the $\sigma$ components of each line will cancel: the quadratic Zeeman effect means that the spacing of various $\sigma$ sub-components is no longer the same in the blue and red features, and line saturation effects will change the equivalent widths of the polarised $\sigma$ components so that they no longer cancel out in the filter measurement. In addition, the filter measurement may be made on a strongly sloped flux continuum, such as those seen in the intensity spectra of the two example stars, so that every blue-wing polarised feature will contribute more photons than the associated red-wing feature. For this reason also, the blue and red polarisation features will not exactly cancel, and the line spectrum will contribute to the net polarisation measured in a broad band that includes lines. 

Thus we expect that polarisation in spectral line features may contribute some net circular polarisation to a broad filter measurement. We note, however, that while this effect means that a particular broad-band measurement may not provide an accurate measurement of the continuum polarisation between spectral lines, the line circular polarisation source still requires the presence of a magnetic field to operate. A broad-band filter measurement of a non-magnetic WD should yield a measurement without significant polarisation, and the detection of a broad-band circular polarisation signal is still a strongly valid sign of the presence of a magnetic field.

\section{Results}
Instrumental polarisation was determined from observations of four bright nearby
stars, assumed to have zero circular polarisation (Sect.~\ref{Sect_Log_Std} and Table~\ref{Tab_Log_Std}). 
The results of our WD observations are given in Table~\ref{Tab_Log_WD} and are presented in Sects.~\ref{Sect_MWDs} to \ref{Sect_Science}.

\begin{table*}
\tabcolsep=0.14cm
  \caption{\label{Tab_Log_WD} Observing log of WDs.  WD\,1900$+$705 and WD\,1831$+$197 were observed twice.}
\begin{center}
\begin{tabular}{lcccrr@{$\pm$}lr@{$\pm$}lr@{$\pm$}l}
\hline\hline
\multicolumn{1}{c}{STAR} &    DATE  &  UT &JD --  & Exp. & \multicolumn{6}{c}{$V/I$ (\%)} \\
                         &yyyy-mm-dd&hh:mm&2400000&\multicolumn{1}{c}{(s)}  &
  \multicolumn{2}{c}{$B'$}&\multicolumn{2}{c}{$V'$} &\multicolumn{2}{c}{$R'$} \\
\hline
WD\,1647$+$591       & 2021-07-03 & 00:59 & 59398.541 & 1920 &$ -0.000$ & 0.005 &$ 0.007$& 0.007 &$ 0.004$& 0.006 \\ 
WD\,2032$+$248       & 2021-07-04 & 05:00 & 59399.709 & 1920 &$  0.001$ & 0.004 &$-0.001$& 0.004 &$-0.002$& 0.005 \\ 
WD\,2117$+$539       & 2021-07-03 & 04:17 & 59398.678 & 1800 &$  0.000$ & 0.006 &$-0.001$& 0.007 &$-0.007$& 0.008 \\[3mm]

{\bf WD\,1748$+$708} & 2021-07-03 & 00:10 & 59398.508 & 1920 &$  0.665$& 0.019 &$ -0.376 $& 0.017 &$-1.126$& 0.014 \\
{\bf WD\,1900$+$705} & 2021-07-02 & 22:22 & 59398.432 &  640 &$  3.756$& 0.016 &$  3.604 $& 0.016 &$ 3.827$& 0.019 \\ [3mm]

WD\,1309$+$853       & 2021-07-02 & 23:10 & 59398.466 & 3120 &$ 0.090 $& 0.048 &$ 0.109$& 0.048&$ 0.113 $& 0.037 \\
WD\,1350$-$090       & 2021-07-03 & 22:03 & 59399.419 & 2240 &$ 0.048 $& 0.015 &$ 0.017$& 0.020&$ 0.011 $& 0.018 \\
WD\,1503$-$070       & 2021-07-03 & 23:17 & 59399.471 & 3360 &$ 0.006 $& 0.030 &$ 0.042$& 0.033&$ 0.061 $& 0.025 \\
WD\,1953$-$011       & 2021-07-03 & 01:48 & 59398.576 & 2240 &$ 0.017 $& 0.012 &$ 0.015$& 0.011&$-0.012 $& 0.014 \\ 
                     & 2021-07-04 & 01:55 & 59399.580 & 3200 &$ 0.016 $& 0.007 &$ 0.002$& 0.008&$ 0.002 $& 0.008 \\
{\bf WD\,2254$+$076} & 2021-07-05 & 04:04 & 59400.671 & 5760 &$-0.347 $& 0.034 &$-0.324$& 0.056&$-0.529 $& 0.055 \\[3mm]

{\bf WD\,1556$+$044} & 2021-07-04 & 22:10 & 59400.424 & 3200 &$-0.218 $& 0.033  &$0.315 $& 0.039 &$-0.131$& 0.030 \\
WD\,1831$+$197       & 2021-07-04 & 00:37 & 59399.526 & 3840 &$-0.003 $& 0.038  &$ 0.049 $& 0.042 &$ 0.087$& 0.037 \\
                     & 2021-07-04 & 23:48 & 59400.493 & 6400 &$ 0.051 $& 0.025  &$-0.015 $& 0.034 &$-0.089$& 0.029 \\
WD\,1856$+$534       & 2021-07-05 & 01:33 & 59400.565 & 3840 &$-0.067 $& 0.058  &$-0.005 $& 0.060 &$-0.032$& 0.043 \\
WD\,2011$+$063       & 2021-07-03 & 02:48 & 59398.618 & 3120 &$ 0.022 $& 0.035  &$0.057 $& 0.040 &$-0.028$& 0.024 \\
{\bf WD\,2049$-$253} & 2021-07-04 & 03:31 & 59399.647 & 6400 &$ 0.442 $& 0.044 &$ 0.460$& 0.039&$ 0.684 $& 0.031 \\ 
{\bf WD\,2049$-$222} & 2021-07-05 & 02:39 & 59400.611 & 2240 &$ 0.095 $& 0.019  &$ 0.068 $& 0.022 &$ 0.142$& 0.020 \\ 
\hline
\end{tabular}
\end{center}
\end{table*}

\subsection{Observations of non-polarised bright stars}\label{Sect_Log_Std}
The high S/N measurements of non-polarised stars yield the instrumental polarisation to a precision better than $10^{-5}$.
In the $B'V'R'$ bands, the values of $V/I$ are 0.0117 $\pm$ 0.0006 \%, 0.0129 $\pm$ 0.0005 \%,
and 0.0084 $\pm$ 0.0004 \%, respectively. The instrumental polarisation was subtracted from the observed
polarisation of all targets, including the measurements of the standard stars in Table\,\ref{Tab_Log_Std}. The WDs in which non-zero circular polarisation was detected by our measurements are highlighted in boldface type in Table~\ref{Tab_Log_WD}. 

\subsection{Known magnetic WDs with continuum polarisation}\label{Sect_MWDs}
\begin{figure}
\centering
\includegraphics[width=8.7cm,trim={0.7cm 5.8cm 1.1cm 3.0cm},clip]{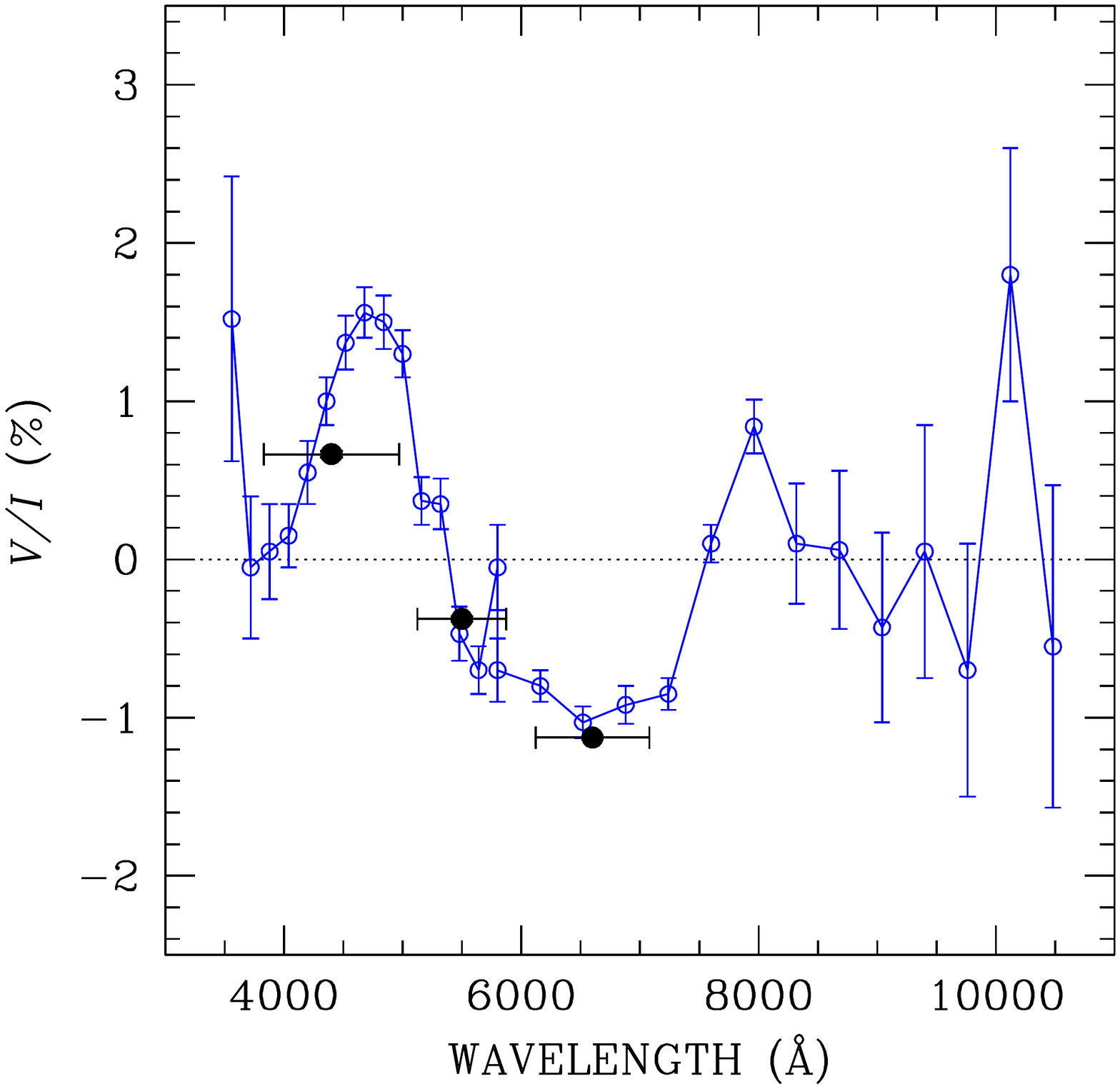}

\includegraphics[width=8.7cm,trim={0.7cm 5.8cm 1.1cm 2.0cm},clip]{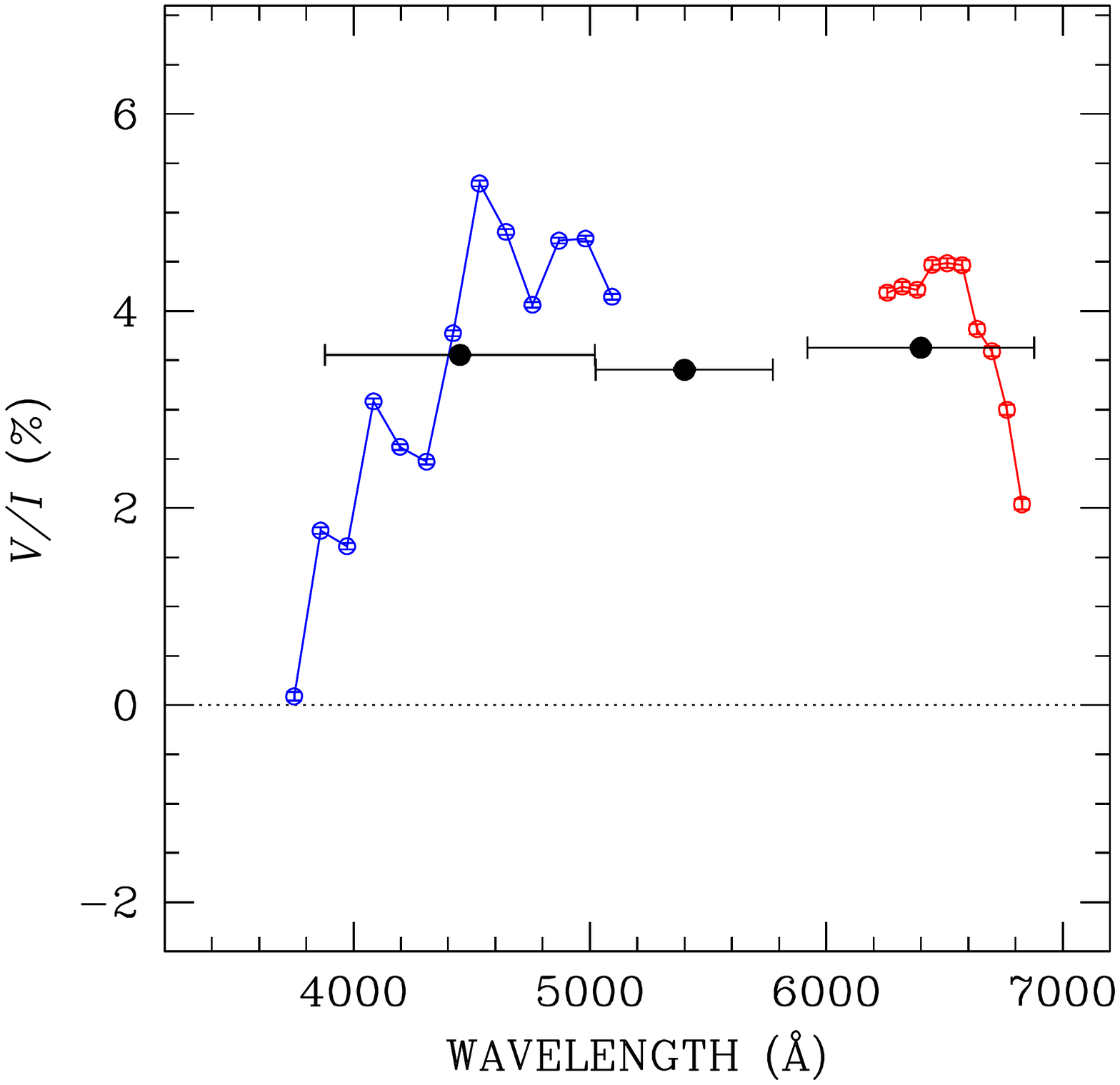}
      \caption{\label{Fig_G240} $V/I$ spectropolarimetry of the continuum (empty symbols joined by solid lines), and new DIPol-UF BBCP (solid circles) of the stars WD\,1748$+$708 (top panel) and WD\,1900$+$705 (bottom panel). Horizontal error bars represent the FWHM of the broad-band filters. The polarisation spectra used for comparison are described in the text. 
}
\end{figure}
WD\,1748$+$708 and WD\,1900$+$705 are well-known high-field MWDs that exhibit strongly polarised continua.  The polarisation of WD\,1748$+$708 changes rapidly with $\lambda$ \citep{Angeetal74}. In Fig.\,2 we compare our filter observations with an unpublished $V/I$ spectrum of WD\,1748+708 obtained by Angel and Landstreet on 1974 September 6.27 UT, using the Oke Multi-Channel Spectrophotometer on the Mount  Palomar 5m telescope, as described for example by \citet{LanAng75}. We note that the sign of $V/I$ reverses within the $V'$ passband. Polarisation values observed in adjacent passbands can be significantly different with broad-band filters. Accordingly, it is essential to make observations in more than one passband  to search efficiently for magnetic WDs.

The spectropolarimetric data shown in the lower panel of Fig.\,2 were obtained with the ISIS instrument on the William Herschel Telescope (WHT). The plotted data are adapted from the detailed search for possible variability of WD\,1900+705 reported by \citet{BagLan19a}. We note that the broad $B'V'R'$ filters faithfully follow the broad outlines of the higher-resolution $V/I$ spectra in both stars, clearly revealing the power of filter polarimetry even for complex polarised spectra. 

\subsection{Non-magnetic WDs}\label{Sect_Non_MWD}
\begin{figure}
\centering
\includegraphics[width=8.7cm,trim={0.7cm 5.8cm 1.1cm 3.0cm},clip]{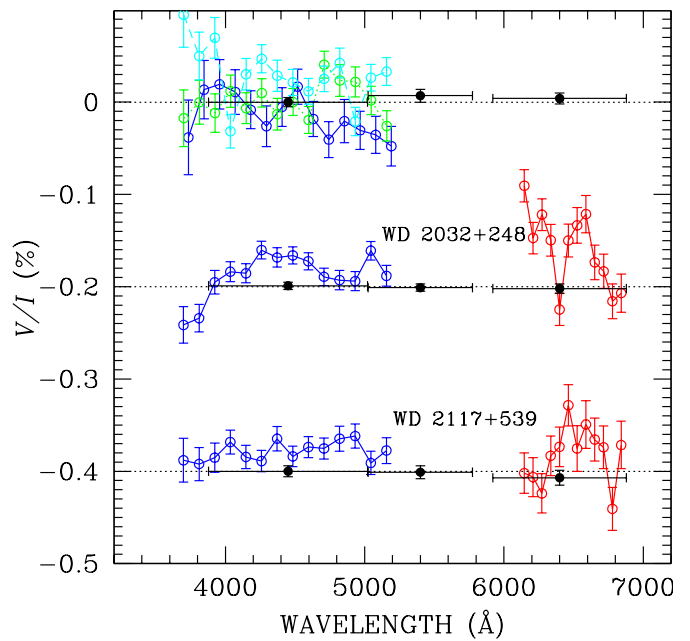}
      \caption{\label{Fig_NMWDs} ISIS spectropolarimetry of the continuum rebinned to 112\,\AA\ in the blue and 63\,\AA\ in the red (empty symbols joined by solid lines), and new DIPol-UF BBCP (solid circles) of three non-magnetic WDs. Star WD\,1647$+$591 was observed three times with grating R600B, and observations obtained in different epochs are shown with different colours. The other two stars were observed simultaneously with gratings R600B and R1200R; see \citet{BagLan18} for more details. Data for different stars have been offset by 0, $-0.2,$ and $-0.4$\,\% for display purposes. Our BBCP measurements (zero polarisation) are consistent with no detectable magnetic field.
}
\end{figure}
Stars WD1647$+$591, WD2032$+$248, and WD2117$+$539 have previously been observed with highly sensitive spectropolarimetric techniques. All were found to be  non-magnetic with sub-kG accuracy through the analysis of the Zeeman effect on the Stokes $V$ profiles of their very strong H Balmer lines. All these stars were observed by \citet{BagLan18} with the ISIS instrument of the WHT. ISIS has the capability of measuring polarisation in the continuum, but it has been found that the precision of these measurements is limited by the difficulty of correcting precisely for spurious background polarisation, particularly in measurements made in the presence of scattered moonlight, as discussed in detail by \citet{BagLan20}. This effect means that it can be difficult to establish the reality of continuum polarisation detected with spectropolarimetry below about the 0.05\,\% level. It is therefore very interesting to compare the very weak continuum polarisation signal observed with ISIS in these three non-magnetic WDs with the results of our new BBCP measurements. This comparison is shown in Fig.~\ref{Fig_NMWDs}. The figure shows that for all non-magnetic WDs from the sample, DiPol-UF has measured zero continuum polarisation ($< 0.01\%$) in  all three passbands. This null result strongly confirms that our BBCP measurements are free of any systematic or calibration errors at the level of $< 10^{-4}$, and it also confirms the earlier conclusion about the serious difficulty of measuring continuum polarisation below about 0.05\,\% with a spectropolarimeter, especially in the presence of moonlit background.

\subsection{Weakly magnetic WDs}\label{Sect_WMWDs}
Our target list included five WDs that have been discovered to be magnetic from the Zeeman splitting of their spectral lines. In increasing order of field strength, these stars are WD\,1350$-$090 and  WD\,1953$-$011, both with $\bz \simeq 0.15$\,MG; 
WD\,1503$-$070, with $\bs \sim 2$\,MG;
WD\,1309$+$853, with $\bs \sim 5.4$\,MG;
and WD\,2254$+$076, with $\bs \sim 16$\,MG. 
These MWDs have been included in the study in order to clarify the relation between known magnetic field properties and observed continuum circular polarisation.

The only star in our sample of weakly magnetic WDs in which we firmly detect a signal of circular polarisation is WD\,2254$+$076, although marginal detections are obtained in  WD\,1309$+$853 and  WD\,1350$-$090. As WD\,2254+076 has the highest \bs\, value of the sample of low-field MWDs, this is not surprising. The other more weakly magnetic stars in this sample probe the limits of our ability to identify weakly magnetic WDs. 

WD\,2254$+$076 was discovered to host a magnetic field based on a flux spectrum from the Sloan Digital Sky Survey (SDSS) showing Zeeman-split H Balmer lines.  \citet{Kuletal09} estimated a field of $\bs \sim 16$\,MG. Assuming that the star has a mean longitudinal field between 0 and 10\,MG, we would expect to measure $\vert V/I \vert$ up to $\sim 0.7$\,\%.  Earlier observations by \citet{Angetal81}, who had measured a BBCP signal of $-0.16 \pm 0.10$\,\%, failed to detect a magnetic field in this star, but our measurements ($-0.53 \pm 0.06$\,\% in $R'$, $-0.32 \pm 0.06$\,\% in $V'$, and  $-0.35 \pm 0.03$\,\% in $B'$) are consistent with our prediction and would serve to definitively identify this WD as magnetic if this were not already known. 

WD\,1309$+$853 = G\,256--7 was discovered to be magnetic by \citet{Putney95}. From the spectra shown in her Fig.~2, we have deduced $\bs \approx 5.4$\,MG, while Putney found 4.9\,MG. The polarisation spectrum reveals a non-zero \bz. From the observations of \citet{Putney95}, we can estimate $\bz \sim 1-3$\,MG. In this star, we have measured $0.113 \pm 0.037$ in the $R'$ filter; in the $B'$ and $V'$ filter we have measured similar values of the polarisation, but only at $\sim 2\,\sigma$ significance level. Because all the Balmer lines in this very cool DA star are extremely weak, all three filter measurements are probably dominated by continuum polarisation, which appears to be roughly constant with wavelength, as is found for the well-known similar DA star WD\,0553--053 = G\,99--47 \citep{AngLan72,Lieetal75}.
Both results are broadly consistent with the estimate that a field for which $\bz = 15$\,MG produces approximately 1\,\% of circular polarisation in the continuum \citep[][and references therein]{BagLan20}, although more and better calibrations are needed to confirm the precision and range of usefulness of this approximation. 

A field characterised by $\bs \sim 2$\,MG was discovered in WD\,1503$-$070 by \citet{Giaetal11} from Zeeman splitting of the shallow and weak Balmer lines. Our measurements set an upper limit of about 0.05\,\% for circular polarisation. From the estimated \bs\ value, we would expect $\vert \bz \vert \la 1$\,MG, that is, a signal of circular polarisation $\la 0.07$\,\%. This estimate is consistent with our measurements. 

In the star WD\,1350$-$090, \citet{SchSmi95} measured $\bz = 85 \pm 9$\,kG. Bagnulo \& Landstreet (unpublished) have measured the value of $\bz$ several times, always with values between +105 and +145\,kG, and \citet{BagLan21} have suggested that this field is actually roughly constant with time. We measured $V/I = 0.048 \pm 0.015$\,\% in the $B'$ filter, but found values lower than 0.02\,\% in the other filters.  For such a small \bz\ value, we would expect a signal of circular polarisation $\la 0.01$\,\%

However, this DA star has strong Balmer lines up to about H8 (where the higher lines merge), and has correspondingly strong line polarisation features throughout the $B'$ filter. Thus we expect that the polarisation features across individual lines may well have failed to cancel exactly, either because of saturation effects, and/or because of the systematically biased weighting given by the continuum slope. Therefore we could reasonably expect some net polarisation precisely in the filter band that contains five Balmer lines, but not in the other two filter bands. In addition, the sharp cutoff of the $B'$ filter is located in the blue wing of H8, and therefore the measurement did not give full weight to the blue $\sigma$ polarisation from this line to cancel the polarised photons from the red $\sigma$ component. Thus the polarisation detected in this band is almost certainly not a measurement of true continuum polarisation, but is a reliable symptom of the presence of magnetic field in this WD. In contrast, the null polarisation  measurements in $V'$ and $R'$ are consistent with the prediction of continuum polarisation weaker than about 0.01\,\%. 

WD\,1953$-$011 has a constant longitudinal field $\la 0.1$\,MG \citep{Valetal08} and should produce a BBCP signal also $\la 0.01$\,\%. Our measurements, which failed to detect circular polarisation, rule out a signal significantly stronger than this value.

\subsection{Science targets}\label{Sect_Science}
We have firmly detected a signal of circular polarisation in three further WDs: WD\,2049$-$253,  WD\,2049$-$222, and  WD\,1556$+$044.
The DC star WD\,2049$-$253 was discovered to be magnetic by \citet{BagLan20}, who measured $V/I \simeq +0.5$\,\%, approximately constant between 370 and 510\,nm,  with the FORS2 instrument in spectropolarimetric mode. The flux spectrum of the star shows no obvious spectral line features, but with an effective temperature slightly below 5000\,K, it could have an atmospheric chemistry dominated by either H or He. Although the spectropolarimetric measurement showed no obvious problems and was obtained during dark time so that polarised sky background due to cross-talk from highly polarised scattered moonlight was not a problem, the first measurement is near the limit of what can reliably be obtained with FORS.  \citet{BagLan20} warned that the discovery needed to be confirmed by further observations. Our new detection of a signal at a similar level to that measured by \citet{BagLan20} fully confirms that the star is indeed magnetic, with a field that is, as estimated by \citet{BagLan20}, probably characterised by $\bz \sim 7$\,MG and so by \bs\ of about 20\,MG or slightly higher.
 
WD\,1556$+$044 and WD\,2049$-$222 are DC WDs that have never been observed in polarimetric mode. They represent the first discovery of new MWDs obtained by means of BBCP since the discoveries in the 1970s that were summarised by \citet{Angetal81}. 

WD\,2049$-$222 was first identified as a DC star in the survey of hot stars in the Galactic halo of \citet{Beretal92}. This classification was confirmed by \citet{KawVen06}, who showed a portion of the spectrum between 6340 and 6780\,\AA. The completely smooth spectrum segment supports the classification of this WD as a DC. There is a significant spread in the values of \teff\ that are found by different investigators, but they generally range between 8000 and 9500\,K. The absence of the slightest indication of H$\alpha$ in the spectrum segment published by \citet{KawVen06} at this effective temperature means that the dominant atmospheric chemistry is very probably He. 
The polarisation signal that we have measured in WD\,2049$-$222 is possibly the weakest non-zero polarisation signal ever detected in a WD. It is weaker than that measured in WD\,0553$+$053 = G99-47 by \citet{AngLan72}, which is $-0.4$\,\% at 4000\,\AA\ and $-0.2$\,\% at 8000\,\AA, and even weaker than that observed in WD\,1116$-$470 by \citet{BagLan21}, which was just $\simeq -0.2$\,\% in the entire range from 4000 to 9000\,\AA, but which in fact still awaits confirmation. 

The broadband polarisation measurements of WD\,2049--222 all have the same sign, but the variations appear to be real, with the maximum signal $V/I = 0.142 \pm 0.020$\,\% found in the $R'$ band. For this star, we estimate $\bz = 1-2$\,MG. The fact that the polarisation in all three broad bands has the same sign is typical of the polarisation spectra of most MWDs showing non-zero continuum circular polarisation, from WDs with very strong continuum polarisation, such as WD\,1900+705 to the weak continuum polarisation of WD\,0553+053. In this star, the available evidence suggests that the observed polarisation is probably true continuum polarisation. 

WD\,1556$+$044 was originally thought to be a quasi-stellar object \citep{VerVer10}, but was recognised as a DC WD by \citet{Limetal13}, who showed its  blue $I$ spectrum to about 5200\,\AA. The spectrum is apparently completely featureless even at high S/N, and with $\teff = 6\,685$\,K, the star very probably has an He-rich atmosphere. Like many MWDs, the mass of this WD, $0.85 M_\odot$, is considerably higher than those of most of the stars in Table\,\ref{Tab_Programme}. The polarisation in WD\,1556+044 is detected at the $6 \sigma$, $8 \sigma,$ and $4 \sigma$ levels in the $B'$, $V',$ and $R'$ filters. In this new MWD, the detection is quite firm, and the polarisation values, which change sign from $B'$ to $V'$ to $R'$,  indicate strong variations of $V/I$ with wavelength. This is an unusual characteristic that is found in only a small minority of MWDs for which the spectrum of $V/I$ is well established. This unusual characteristic is perhaps most similar to the variation seen in WD\,1748+708 = G240-72 (Fig.\,\ref{Fig_G240}), which is only about 1\,000\,K cooler. The explanation for the unusual sign changes in these two MWDs is unknown, but may involve similar chemical pollution of the nearly pure He atmospheres. This new MWD could have a magnetic field of $\bz \sim 5$\,MG, and a field \bs\ that is at least 12 or 15\,MG, and perhaps larger.

We now comment on the remaining three science targets, for which we are able to set only upper limits for $\vert V/I \vert$ and $\vert \bz \vert$. Among these targets,  WD\,1856$+$534 is of special interest because \citet{Vanetal20} discovered a giant planet orbiting it that has not been tidally disrupted. The star has never been observed in polarimetric mode, and our measurements suggest that the star's circular polarisation is most likely $< 0.1$\,\% (in absolute value), setting our estimate of the upper limit for $\vert \bz\ \vert$ to 1MG. 

We observed WD\,1831$+$197 twice. With our first observation, we obtained a detection of nearly 3\,$\sigma$  with the $R'$ filter. Taken at a face value, our second observation represents a 3\,$\sigma$ detection (but with the opposite sign with respect to the previous night). Owing to the lack of detection in the other filters, we cannot yet accept the star as a MWD, but it would be worthwhile to re-observe it with a larger telescope, or with a much longer exposure time. The star could have a field of about 1\,MG and might be variable. At the same time, our observations probably rule out the presence of a field with a longitudinal component larger than a few MG.

WD\,2011$+$063 was one of the first WDs ever observed in polarimetric mode: \citet{AngLan70} measured $V/I = 0.12 \pm 0.09$\,\%. Our new measurements are about three times more accurate, but still detect no polarisation signal. The upper limit for $\vert \bz\ \vert$ is again of about 1\,MG.

\section{Polarimetric sensitivity and detection limits} 
\begin{figure}
\centering
\includegraphics[width=8.7cm,trim={0.7cm 5.8cm 1.1cm 3.0cm},clip]{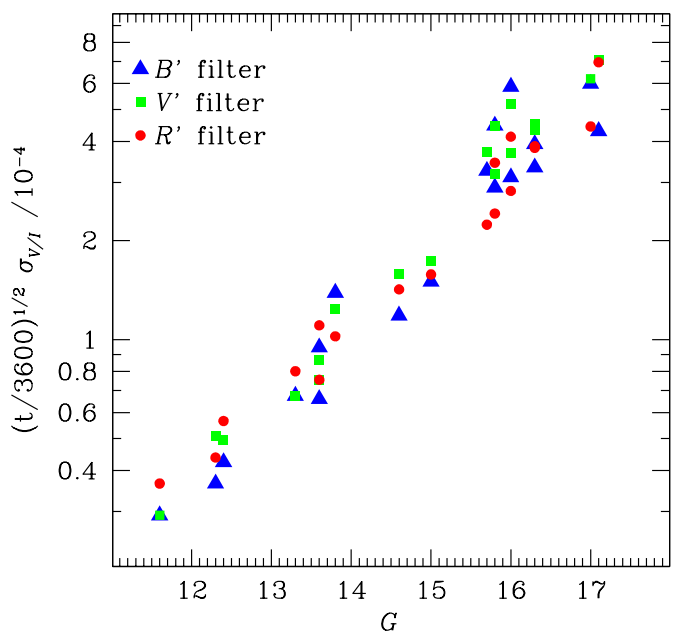}
      \caption{\label{Fig_ETC} Relation between magnitude and uncertainty of our measurements of Table~\ref{Tab_Log_WD} as a function of Gaia $G$ magnitude, normalised to 1\,h exposure time.} 
\end{figure}
Our observing run may be used to make quantitative predictions regarding the level of polarimetric sensitivity that may be achieved for the BBCP measurements as a function of time, stellar magnitude, and telescope size  using DIPol-UF with our filter set. Fig.~\ref{Fig_ETC} shows our measurement uncertainties in the $B'V'R'$ filters, normalised to 1\,h exposure time, as a function of the Gaia magnitude $G$. Uncertainties depend on the spectral energy distribution of the star, that is, for hot stars, the increased flux in the blue reduces errors in the $B'$ band, whereas for cool stars, the best S/N is achieved in the red. For our sample of stars and their spectral energy dependence, the errors are rather uniformly distributed (Fig.~\ref{Fig_ETC}). The high throughput of the $B'$ band (Fig.1) helps to balance the performance of the DIPol-UF instrument in the different passbands recorded simultaneously. 

The relationship obtained for the polarimetric uncertainty as a function of the input flux indicates that the former is defined by the photon flux with no detectable systematic errors (both axes in Fig.~\ref{Fig_ETC} are logarithmic). This nicely demonstrates that the DIPol-UF polarimeter provides polarimetric measurements with precision and accuracy down to 10 parts per million (ppm) also in circular polarisation, that is, it is comparable to the few ppm uncertainty achieved in linear polarisation measurements \citep{Piietal21}. Thus, using this relation, we can evaluate observational requirements for a given WD magnetic field with DIPol-UF employed at various telescopes.

From the estimate that a $\bz = 15$\,MG field produces a polarisation of approximately 1\,\% in the continuum \citep[][and references therein]{BagLan20}, it follows that a signal of $\sim$ 0.07 \% is expected from a $\bz = 1$\,MG field. A 3$\sigma$ detection of this requires 
an uncertainty of $\sim$ 0.02\%. According to the relation in Fig. 4, this is 
achieved in one hour for a star with a magnitude of $G \sim 15.5$. For a brighter 
$G \sim 12$ star, the precision 0.004\% in one hour would yield a 3$\sigma$ 
detection of $\bz \sim 0.2$\,MG. For the faintest ($G \sim $17) stars in our sample, the precision achieved in one hour would allow $> 3 \sigma$ detection of fields $\bz > 3$\,MG. 

The typical seeing values (0.6" - 1.0") and the sky darkness during our observations are comparable to the conditions at the best observing sites, and our detection limits could easily be scaled to larger telescopes. For an 8 m telescope, the measured polarisation uncertainties would be lower by a factor of $\sim 3$ for a given exposure time.

\section{Conclusions}
During the past two years, five new MWDs of spectral class DC have been discovered: WD\,0004$+$122, WD\,078$-$670, WD\,2049$-$253 \citep[discoveries presented by][with confirmation for the latter obtained in this work]{BagLan20},  WD\,1556$+$044, and WD\,2049$-$222 (this work), while the discovery of a $\sim 20$\,MG field in WD\,1116$-$47 by \citet{BagLan21} still needs to be confirmed. These are the first discoveries of magnetic fields in old, featureless WDs since the mid-1990s by \citet{Putney95}. WD\,1556$+$044 and WD\,2049$-$222 represent the first discoveries obtained by means of BBCP filter measurements since the late 1970s. The significance is that we are now able to explore the magnetic features of stars in the last stages of stellar evolution. In particular, studying the frequency of the occurrence of magnetic fields in old WDs will allow us to understand if and when Ohmic decay acts during the long WD cooling phase, or if magnetic fields are found with high frequency and high field strength even among the very oldest WDs of our neighbourhood.

\begin{acknowledgements}
Based on observations made with the Nordic Optical Telescope, owned in collaboration by the University of Turku and Aarhus University,
and operated jointly by Aarhus University, the University of Turku and the University of Oslo, representing Denmark, Finland and Norway, the University of Iceland and Stockholm University at the Observatorio del Roque de los Muchachos, La Palma, Spain, of the Instituto de Astrofisica de Canarias. DIPol-UF is a joint effort between University of Turku (Finland) and Leibniz Institute for Solar Physics (Germany). We  acknowledge  support from the Magnus Ehrnrooth foundation and ERC Advanced Grant Hot-Mol  ERC-2011-AdG-291659.
JDL acknowledges the financial support of the Natural Sciences and Engineering Research Council of Canada (NSERC), funding reference number 6377-2016.
\end{acknowledgements}
\bibliographystyle{aa}
\bibliography{sbabib}

\begin{thebibliography}{38}
\expandafter\ifx\csname natexlab\endcsname\relax\def\natexlab#1{#1}\fi

\bibitem[{{Angel} {et~al.}(1981){Angel}, {Borra}, \& {Landstreet}}]{Angetal81}
{Angel}, J.~R.~P., {Borra}, E.~F., \& {Landstreet}, J.~D. 1981, \apjs, 45, 457

\bibitem[{{Angel} {et~al.}(1974{\natexlab{a}}){Angel}, {Hintzen},
  {Strittmatter}, \& {Martin}}]{Angetal74}
{Angel}, J.~R.~P., {Hintzen}, P., {Strittmatter}, P.~A., \& {Martin}, P.~G.
  1974{\natexlab{a}}, \apjl, 190, L71

\bibitem[{{Angel} {et~al.}(1974{\natexlab{b}}){Angel}, {Hintzen},
  {Strittmatter}, \& {Martin}}]{Angeetal74}
{Angel}, J.~R.~P., {Hintzen}, P., {Strittmatter}, P.~A., \& {Martin}, P.~G.
  1974{\natexlab{b}}, \apjl, 190, L71

\bibitem[{{Angel} \& {Landstreet}(1970{\natexlab{a}})}]{AngLan70}
{Angel}, J.~R.~P. \& {Landstreet}, J.~D. 1970{\natexlab{a}}, \apjl, 162, L61

\bibitem[{{Angel} \& {Landstreet}(1970{\natexlab{b}})}]{AngLan70a}
{Angel}, J.~R.~P. \& {Landstreet}, J.~D. 1970{\natexlab{b}}, \apjl, 160, L147

\bibitem[{{Angel} \& {Landstreet}(1972)}]{AngLan72}
{Angel}, J.~R.~P. \& {Landstreet}, J.~D. 1972, \apjl, 178, L21

\bibitem[{{Aznar Cuadrado} {et~al.}(2004){Aznar Cuadrado}, {Jordan},
  {Napiwotzki}, {Schmid}, K, \& G.}]{Aznar2004}
{Aznar Cuadrado}, R., {Jordan}, S., {Napiwotzki}, R., {et~al.} 2004, \aap, 423,
  1081

\bibitem[{{Bagnulo} \& {Landstreet}(2018)}]{BagLan18}
{Bagnulo}, S. \& {Landstreet}, J.~D. 2018, \aap, 618, A113

\bibitem[{{Bagnulo} \& {Landstreet}(2019)}]{BagLan19a}
{Bagnulo}, S. \& {Landstreet}, J.~D. 2019, \mnras, 486, 4655

\bibitem[{{Bagnulo} \& {Landstreet}(2020)}]{BagLan20}
{Bagnulo}, S. \& {Landstreet}, J.~D. 2020, \aap, 643, A134

\bibitem[{{Bagnulo} \& {Landstreet}(2021)}]{BagLan21}
{Bagnulo}, S. \& {Landstreet}, J.~D. 2021, \mnras, 507, 5902

\bibitem[{{Berdyugin} \& {Piirola}(1999)}]{BerPii99}
{Berdyugin}, A.~V. \& {Piirola}, V. 1999, \aap, 352, 619

\bibitem[{{Bergeron} {et~al.}(1992){Bergeron}, {Ruiz}, \&
  {Leggett}}]{Beretal92}
{Bergeron}, P., {Ruiz}, M.-T., \& {Leggett}, S.~K. 1992, \apj, 400, 315

\bibitem[{{Gianninas} {et~al.}(2011){Gianninas}, {Bergeron}, \&
  {Ruiz}}]{Giaetal11}
{Gianninas}, A., {Bergeron}, P., \& {Ruiz}, M.~T. 2011, \apj, 743, 138

\bibitem[{{Isern} {et~al.}(2017){Isern}, {Garsia-Berro}, {Kulebi}, \&
  {Loren-Aguilar}}]{Isern2017}
{Isern}, J., {Garsia-Berro}, E., {Kulebi}, B., \& {Loren-Aguilar}, P. 2017,
  \apjl, 836, L28

\bibitem[{{Kawka} \& {Vennes}(2006)}]{KawVen06}
{Kawka}, A. \& {Vennes}, S. 2006, \apj, 643, 402

\bibitem[{{Kemp} {et~al.}(1970){Kemp}, {Swedlund}, {Landstreet}, \&
  {Angel}}]{Kemetal70}
{Kemp}, J.~C., {Swedlund}, J.~B., {Landstreet}, J.~D., \& {Angel}, J.~R.~P.
  1970, \apjl, 161, L77

\bibitem[{{K{\"u}lebi} {et~al.}(2009){K{\"u}lebi}, {Jordan}, {Euchner},
  {G{\"a}nsicke}, \& {Hirsch}}]{Kuletal09}
{K{\"u}lebi}, B., {Jordan}, S., {Euchner}, F., {G{\"a}nsicke}, B.~T., \&
  {Hirsch}, H. 2009, \aap, 506, 1341

\bibitem[{{Landstreet} \& {Angel}(1971)}]{LanAng71}
{Landstreet}, J.~D. \& {Angel}, J.~R.~P. 1971, \apjl, 165, L67

\bibitem[{{Landstreet} \& {Angel}(1975)}]{LanAng75}
{Landstreet}, J.~D. \& {Angel}, J.~R.~P. 1975, \apj, 196, 819

\bibitem[{{Landstreet} \& {Bagnulo}(2019)}]{LanBag19a}
{Landstreet}, J.~D. \& {Bagnulo}, S. 2019, \aap, 623, A46

\bibitem[{{Liebert} {et~al.}(1975){Liebert}, {Angel}, \&
  {Landstreet}}]{Lieetal75}
{Liebert}, J., {Angel}, J.~R.~P., \& {Landstreet}, J.~D. 1975, \apjl, 202, L139

\bibitem[{{Liebert} \& {Stockman}(1980)}]{LieSto80}
{Liebert}, J. \& {Stockman}, H.~S. 1980, \pasp, 92, 657

\bibitem[{{Limoges} {et~al.}(2013){Limoges}, {L{\'e}pine}, \&
  {Bergeron}}]{Limetal13}
{Limoges}, M.~M., {L{\'e}pine}, S., \& {Bergeron}, P. 2013, \aj, 145, 136

\bibitem[{{Patat} \& {Romaniello}(2006)}]{patat2006}
{Patat}, F. \& {Romaniello}, M. 2006, \pasp, 118, 146

\bibitem[{{Piirola} {et~al.}(2014){Piirola}, {Berdyugin}, \&
  {Berdyugina}}]{Piietal14}
{Piirola}, V., {Berdyugin}, A., \& {Berdyugina}, S. 2014, in Society of
  Photo-Optical Instrumentation Engineers (SPIE) Conference Series, Vol. 9147,
  Ground-based and Airborne Instrumentation for Astronomy V, ed. S.~K.
  {Ramsay}, I.~S. {McLean}, \& H.~{Takami}, 91478I

\bibitem[{{Piirola} {et~al.}(2021){Piirola}, {Kosenkov}, {Berdyugin},
  {Berdyugina}, \& {Poutanen}}]{Piietal21}
{Piirola}, V., {Kosenkov}, I.~A., {Berdyugin}, A.~V., {Berdyugina}, S.~V., \&
  {Poutanen}, J. 2021, \aj, 161, 20

\bibitem[{{Piirola, V.} {et~al.}(2020){Piirola, V.}, {Berdyugin, A.}, {Frisch,
  P. C.}, {Kagitani, M.}, {Sakanoi, T.}, {Berdyugina, S.}, {Cole, A. A.},
  {Harlingten, C.}, \& {Hill, K.}}]{piirola2020a}
{Piirola, V.}, {Berdyugin, A.}, {Frisch, P. C.}, {et~al.} 2020, A\&A, 635, A46

\bibitem[{{Putney}(1995)}]{Putney95}
{Putney}, A. 1995, \apjl, 451, L67

\bibitem[{{Putney}(1997)}]{Putney97}
{Putney}, A. 1997, \apjs, 112, 527

\bibitem[{{Putney} \& {Jordan}(1995)}]{PutJor95}
{Putney}, A. \& {Jordan}, S. 1995, \apj, 449, 863

\bibitem[{{Schmidt} \& {Smith}(1994)}]{SchSmi94}
{Schmidt}, G.~D. \& {Smith}, P.~S. 1994, \apjl, 423, L63

\bibitem[{{Schmidt} \& {Smith}(1995)}]{SchSmi95}
{Schmidt}, G.~D. \& {Smith}, P.~S. 1995, \apj, 448, 305

\bibitem[{{Shipman}(1971)}]{Ship71}
{Shipman}, H.~L. 1971, \apj, 167, 165

\bibitem[{{Valyavin} {et~al.}(2008){Valyavin}, {Wade}, {Bagnulo}, {Szeifert},
  {Landstreet}, {Han}, \& {Burenkov}}]{Valetal08}
{Valyavin}, G., {Wade}, G.~A., {Bagnulo}, S., {et~al.} 2008, \apj, 683, 466

\bibitem[{{Vanderburg} {et~al.}(2020){Vanderburg}, {Rappaport}, {Xu},
  {Crossfield}, {Becker}, {Gary}, {Murgas}, {Blouin}, {Kaye}, {Palle}, {Melis},
  {Morris}, {Kreidberg}, {Gorjian}, {Morley}, {Mann}, {Parviainen}, {Pearce},
  {Newton}, {Carrillo}, {Zuckerman}, {Nelson}, {Zeimann}, {Brown},
  {Tronsgaard}, {Klein}, {Ricker}, {Vanderspek}, {Latham}, {Seager}, {Winn},
  {Jenkins}, {Adams}, {Benneke}, {Berardo}, {Buchhave}, {Caldwell},
  {Christiansen}, {Collins}, {Col{\'o}n}, {Daylan}, {Doty}, {Doyle},
  {Dragomir}, {Dressing}, {Dufour}, {Fukui}, {Glidden}, {Guerrero}, {Guo},
  {Heng}, {Henriksen}, {Huang}, {Kaltenegger}, {Kane}, {Lewis}, {Lissauer},
  {Morales}, {Narita}, {Pepper}, {Rose}, {Smith}, {Stassun}, \&
  {Yu}}]{Vanetal20}
{Vanderburg}, A., {Rappaport}, S.~A., {Xu}, S., {et~al.} 2020, \nat, 585, 363

\bibitem[{{V{\'e}ron-Cetty} \& {V{\'e}ron}(2010)}]{VerVer10}
{V{\'e}ron-Cetty}, M.~P. \& {V{\'e}ron}, P. 2010, \aap, 518, A10

\bibitem[{{West}(1989)}]{West89}
{West}, S.~C. 1989, \apj, 345, 511

\end{thebibliography}

\end{document}